\documentclass[fleqn,usenatbib]{mnras}
\usepackage[T1]{fontenc}
\DeclareRobustCommand{\VAN}[3]{#2}
\let\VANthebibliography\thebibliography
\def\thebibliography{\DeclareRobustCommand{\VAN}[3]{##3}\VANthebibliography}

\usepackage{graphicx}	
\usepackage{amsmath}	
\usepackage{amssymb}	
\usepackage{siunitx,fancyvrb,xcolor,bm}
\DeclareSIUnit\parsec{pc}

\newcommand{\ud}{\mathrm{d}}    
\newcommand{\ue}{\mathrm{e}}    

\def\lya{Lyman-$\alpha$}
\hypersetup{unicode, breaklinks}
\usepackage[utf8]{inputenc}

\newcommand{\lymana}{{Lyman-\ensuremath{\upalpha}}}
\newcommand{\lymanatext}{Lyman-α}
\usepackage{upgreek}

\defcitealias{Hirata}{H06}
\defcitealias{Meiksin}{M06}
\defcitealias{Chen}{CM04}
\defcitealias{Chuzhoy_2007}{CS07}
\defcitealias{Ghara}{GM19}
\defcitealias{FP06}{FP06}
\title[Ly~$\alpha$ coupling and heating]{\texorpdfstring{\lymana}{\lymanatext} coupling and heating at Cosmic Dawn}

\author[S. Mittal and G. Kulkarni]{
Shikhar Mittal\thanks{E-mail: shikhar.mittal@tifr.res.in}
and Girish Kulkarni\thanks{E-mail: kulkarni@theory.tifr.res.in}
\\
Tata Institute of Fundamental Research, Homi Bhabha Road, Mumbai 400005, India
}

\date{Accepted 2020 December 7. Received 2020 December 2; in original form 2020 September 23}

\pubyear{2020}

\begin{document}
\label{firstpage}
\pagerange{\pageref{firstpage}--\pageref{lastpage}}
\maketitle

\begin{abstract}
The global 21-cm signal from the cosmic dawn is affected by a variety of heating and cooling processes.  We investigate the impact of heating due to Lyman-$\alpha$ (Ly~$\alpha$) photons on the global 21-cm signal at cosmic dawn using an analytical expression of the spectrum around the Ly~$\alpha$ resonance based on the so-called `wing approximation'. We derive a new expression for the scattering correction and for the first time give a simple close-form expression for the cooling due to injected Ly~$\alpha$ photons. We perform a short parameter study by varying the Ly~$\alpha$ background intensity by four orders of magnitude and establish that a strong Ly~$\alpha$ background is necessary, although not sufficient, in order to reproduce the recently detected stronger-than-expected 21-cm signal by the EDGES Collaboration. We show that the magnitude of this Ly~$\alpha$ heating is smaller than previously estimated in the literature by two orders of magnitude or more.  As a result, even a strong Ly~$\alpha$ background is consistent with the EDGES measurement. We also provide a detailed discussion on different expressions of the Ly~$\alpha$ heating rate used in the literature.
\end{abstract}



\section{Introduction}

The 21-cm signal arising due to the hyperfine splitting of ground state of the neutral hydrogen atom is a promising probe of the cosmic dawn and epoch of reionization (EoR) \citep{MMR}. The interaction of the magnetic dipole moment of the proton and that of the electron splits the ground state of the hydrogen atom into two levels separated by a small energy of $\Delta E=hc/\lambda_{21}$, where $\lambda_{21}=\SI{0.21}{\metre}$ \citep{woodgate}. Cosmology using the 21-cm line is reviewed extensively by \citet{Bar, Furlanetto, Pritchard_2012, BARKANA_2016}.

The strength of the 21-cm signal depends on various astrophysical and cosmological processes, many of which are poorly understood. It captures the thermal and ionization state of the Universe, which in turn are a probe of the formation of first stars \citep{Barkana18, Mesinger19}. An important process affecting the 21-cm signal is the Wouthuysen--Field (WF) effect \citep{Field, Wouth}. This refers to a change in the occupation numbers of hyperfine states due to resonance scattering of \lya\ (Ly~$\alpha$) photons by the hydrogen atom. This effect makes the 21-cm signal distinguishable from the cosmic microwave background (CMB).
 
Investigation into the physics of the global 21-cm cosmological signal has recently been re-energised due to detection of a signal at redshift $z\sim 17$ by the Experiment to Detect the Global EoR Signal (EDGES) collaboration \citep{Bowman}. The detection reported has an amplitude that is more than double that predicted by the most optimistic theoretical models. While the cosmological nature of this signal is still being investigated \citep{Hills, Bradley, Singh_2019, Sims}, there have been many new exciting theories which try to explain its anomalous amplitude. Broadly speaking there are two types of ideas in the literature. The first type considers lower matter temperature than the estimates of adiabatic cooling \citep{Bar18, Berlin, Munoz, Liu}. The second type considers an excess radio background above the CMB \citep{Feng_2018, Ewall, Fialkov, Ewall2}. The end result of both hypotheses is an increase in the amplitude of the predicted 21-cm absorption signal.

Several groups are working to validate the EDGES claim. Some projects that are already active or under development are the Large Aperture Experiment to Detect the Dark Ages \citep[LEDA,][]{Bernardi_2015, Bernardi_16, Price}, the Shaped Antenna measurement of the background RAdio Spectrum \citep[SARAS,][]{Patra, Singh_2017}, Probing Radio Intensity at high-Z from Marion \citep[PRIzM,][]{philip}, Radio Experiment for the Analysis of Cosmic Hydrogen\footnote{\url{https://www.kicc.cam.ac.uk/projects/reach}} \citep[REACH,][]{REACH}, Sonda Cosmol{\'o}gica de las Islas para la Detecci{\'o}n de Hidr{\'o}geno Neutro \citep[SCI-HI,][]{Scihi} and the Cosmic Twilight Polarimeter \citep[CTP,][]{Nhan_2017, Nhan_2019}.
 
In this paper we reconsider the effect of Ly~$\alpha$ photons on the global 21-cm cosmological signal.  We compute the amount of scattering and heating expected from Ly~$\alpha$ photons at cosmic dawn.  In the process, we derive a new expression for the scattering correction and give a simple closed form expression for the cooling part due to the injected Ly~$\alpha$ photons.  In order to understand constraints on the high-redshift Ly~$\alpha$ background due to the EDGES measurement, we perform a simple single-parameter study where we vary the strength of Ly~$\alpha$ radiation background by 4 orders of magnitude to see its effect on the temperature of intergalactic medium (IGM) and correspondingly on the differential brightness temperature.  Because our purpose here is to gauge the effects of Ly~$\alpha$ radiation only, we do not include processes such as X-ray heating \citep[for e.g.][]{Mesinger12}, shock heating \citep{F04}, etc. The redshift range of our interest is $14\leqslant z\leqslant30$.

Ly~$\alpha$ scattering and heating have been considered in the literature before. \citet{Field} presented the earliest treatment, in which there was no Ly~$\alpha$ heating since it was assumed there are no spectral distortions in the Ly~$\alpha$ spectrum. \citet{MMR} improved this by accounting for the latter but considered the hydrogen atoms to be at rest, which overestimated the Ly~$\alpha$ heating. The first major improvement in the problem came from \citet[hereafter \citetalias{Chen}]{Chen} who solved the radiative transfer equation numerically under the diffusion approximation (Fokker--Planck equation). \citet[hereafter \citetalias{FP06}]{FP06} gave analytical estimates using the analytical solutions of \citet{Chuzhoy} based on a further approximation called the wing approximation. Recently \citet[hereafter \citetalias{Ghara}]{Ghara} also studied Ly~$\alpha$ heating at cosmic dawn and argued that a strong Ly~$\alpha$ background radiation is ruled out in view of the EDGES claim. 

This paper is organized as follows. In Section~\ref{TM} we present the theory of scattering and heating by Ly~$\alpha$ photons and their effect on the 21-cm signal. In Section~\ref{RA} we present our results and analysis.  We discuss our conclusions and ideas on further work in Section~\ref{Con}. The following cosmological parameters are used: $\Omega_\mathrm{m}= 0.32, \Omega_\mathrm{b}=0.049, \Omega_\Lambda = 0.68, h=0.67, Y_{\mathrm{p}}=0.24, T_0=\SI{2.73}{\kelvin}, \sigma_8 = 0.83$ and $n_{\mathrm{s}} = 0.96$ \citep{Planck}, where $T_0$ and $Y_\mathrm{p}$ are the CMB temperature measured today and the helium fraction by mass, respectively. Unless stated otherwise, we will work in SI units. The reader is cautioned here as they may find some of our expressions differing from those in previous literature, which use CGS units, by a factor of $4\pi\varepsilon_0$.

\section{Theory and Methods}\label{TM}

We begin by writing down the observable 21-cm signal, which is the 21-cm brightness temperature measured against the CMB temperature and is given by \citep{F06}
\begin{multline}
\Delta T_\mathrm{b}=27\bar{x}_{\ion{H}{i}}\left(\frac{1-Y_{\mathrm{p}}}{0.76}\right)\left(\frac{\Omega_\mathrm{b}h^2}{0.023}\right)\sqrt{\frac{0.15}{\Omega_\mathrm{m}h^2}\frac{1+z}{10}}\\\times\left(1-\frac{T_\gamma}{T_\mathrm{s}}\right)\si{\milli\kelvin}\,,\label{DeltaT}
\end{multline} 
where $T_\mathrm{s}$ is the spin temperature, $T_\gamma$ is CMB temperature, $z$ is the redshift, $x_{\ion{H}{i}}\equiv n_{\ion{H}{i}}/n_{\text{H}}$ is the ratio of number densities of neutral hydrogen (\ion{H}{i}) and total hydrogen (H), and we have assumed a matter dominated Universe, so that
$H(z)=H_0\sqrt{\Omega_\mathrm{m}(1+z)^3}$. The signal is seen in absorption when $\Delta T_\mathrm{b}<0$ and when $\Delta T_\mathrm{b}>0$ the signal is seen in emission. (Note that in equation~\ref{DeltaT} we write $h$ to represent the Hubble's constant in units of $\SI{100}{\kilo\metre\per\second\per\mega\parsec}$ for the last time. From here onwards $h$ will denote Planck's constant.)

Before reionization, the globally volume averaged neutral hydrogen fraction is same as that measured in the bulk of IGM, so that $\bar{x}_{\ion{H}{i}}=x_{\ion{H}{i}}$. Moreover, neglecting the electron contribution from helium, it is safe to write
\begin{equation}
x_{\ion{H}{i}}=1-x_{\text{e}}\,,
\end{equation}
where $x_{\text{e}}\equiv n_{\text{e}}/n_{\text{H}}$ is ratio of number of electrons to the total number of hydrogen atoms. The $x_{\text{e}}$ values may be obtained for our cosmology at high redshifts using \textsc{recfast}\footnote{\url{https://www.astro.ubc.ca/people/scott/recfast.html}.} \citep{Seager_1999}. 

The spin temperature is defined as the temperature required to achieve a given ratio of populations of upper and lower hyperfine levels, i.e.,
\begin{equation}
\frac{n_1}{n_0}\equiv3\ue^{-T_*/T_\mathrm{s}}\,,
\end{equation}
where $T_{*}= h\nu_{21}/k_\mathrm{B}=\SI{0.068}{\kelvin}$, $h$ is the Planck's constant, $k_\mathrm{B}$ is the Boltzmann constant and $\nu_{21}=\SI{1420}{\mega\hertz}$. The factor of 3 is due to the degeneracy factor. The interaction of \ion{H}{i} with the CMB photons, collisions with the other hydrogen atoms/electrons, and the interaction of Ly~$\alpha$ photons determine $T_\mathrm{s}$ \citep{Field}. As a result, it can be expressed as a weighted arithmetic mean of $T_\gamma, T_\mathrm{K}$ and $T_\alpha$ which represent the CMB temperature, gas kinetic temperature and colour temperature, respectively \citep{F06}. Thus,
\begin{equation}
T_{\mathrm{s}}^{-1}=\frac{T_\gamma^{-1}+ x_{\mathrm{K}}T_\mathrm{K}^{-1}+ x_\alpha T_\alpha^{-1}}{1+x_{\mathrm{K}}+x_\alpha}\,.\label{Ts}
\end{equation}
where the collisional coupling and Ly~$\alpha$ coupling are 
\begin{align}
x_\mathrm{K}=\frac{T_*C_{10}}{T_\gamma A_{21}}\,,\label{xk}\\
x_\alpha=\frac{T_*P_{10}}{T_\gamma A_{21}}\,,\label{xa}
\end{align}
respectively. Here, $C_{10}$ and $P_{10}$ are the de-excitation rates by collisions and Ly~$\alpha$ photons, respectively, and $A_{21}=\SI{2.85e-15}{\hertz}$ is the Einstein coefficient of spontaneous emission for the hyperfine transition. The collisional de-excitation rate $C_{10}$ can be written as $n_{\text{e}}\kappa_{\text{eH}}+n_{\ion{H}{i}}\kappa_{\text{HH}}$, where $n_{i}$ is the number density of species $i$ and $\kappa_{ij}$ is the specific rate coefficient in units of volume per unit time. Expressions for $\kappa$s as a function of temperature can be found in \citet{Liszt}. The final expression for $x_\mathrm{K}$ is
\begin{equation}
x_{\mathrm{K}}(z)=\frac{T_*n_\text{H}}{T_\gamma A_{21}}[x_{\text{e}}\kappa_{\text{eH}}+(1-x_{\text{e}})\kappa_{\text{HH}}]\,.\label{XK2}
\end{equation}

The variation of gas kinetic temperature with redshift is important in the understanding of the 21-cm signal. For this we use the following thermal equation
\begin{equation}
(1+z)\frac{\ud T_\mathrm{K}}{\ud z}=2T_\mathrm{K}-\frac{T_{\mathrm{K}}(1+z)}{1+x_{\text{He}}+x_\text{e}}\frac{\ud x_\text{e}}{\ud z}-\frac{2}{3n_{\mathrm{b}}k_\mathrm{B}H}\sum q\,,\label{DEofTk}
\end{equation}
where $n_{\mathrm{b}}=n_{\text{H}}(1+x_{\text{He}}+x_{\text{e}})$ is the total particle number density and similar to $x_{\text{e}}$, we can define $x_\text{He}$. For the given cosmological parameter $Y_{\mathrm{p}}$ it is
\begin{equation}
x_\text{He}=\frac{Y_{\mathrm{p}}}{4(1-Y_{\mathrm{p}})}\,.
\end{equation}
For simplicity we have ignored the electron contribution from helium. There are different processes which heat (or cool) the IGM. To account for this we have the last term on right hand side of equation~\eqref{DEofTk}, in which the summation implies addition of all heating rates denoted by $q$. Note that both $q$ and $n_\text{H}$ should either be in comoving units or both in proper units. Equation~\eqref{DEofTk} is the most general form of thermal evolution. However, for redshifts of our interest the second term on the right hand side may be dropped since the electron fraction changes negligibly and because $x_\text{e}$ itself is quite small ($\sim\num{e-4}$), Compton heating may be neglected \citep{Seager_1999, Seager_2000, Haimoud}. Thus, in this work we will consider only the Ly~$\alpha$ heating term.  We will integrate equation~\eqref{DEofTk} from $z=30$ to $z=14$ with the initial condition, obtained from \textsc{recfast}, $T_{\mathrm{K}}(z=30)=\SI{18}{\kelvin}$.

\subsection{\texorpdfstring{Effect of Scattering of Ly~$\alpha$ Photons}{Effect of Scattering of Ly~α Photons}}

As galaxy formation begins in the Universe at $z\lesssim50$ \citep{Naoz}, ultraviolet (UV) photons produced by the stars in these galaxies are injected into the IGM.  Of particular interest in this work are the Ly~$\alpha$ photons. They play a dual role in the physics of the 21-cm signal.  First, the Ly~$\alpha$ photons couple the 21-cm spin temperature to the gas kinetic temperature. Second, they also change the gas temperature, usually heating the gas.  We discuss the coupling in this section.

An excitation from the ground state followed by a de-excitation due to scattering of UV photons can cause hyperfine transitions in \ion{H}{i}. A hydrogen atom in the first excited state may return to a different hyperfine state it originally started from. The photons so involved are from the Lyman series. This effect is called the Wouthuysen--Field effect \citep{Wouth, Field}. Naturally, there is some energy exchange in this process between the two species and as a result the system tends to achieve an equilibrium. The `heat reservoir' of the Ly~$\alpha$ photons can be given an artificial temperature called the colour temperature, $T_\alpha$ \citep{MMR}. 

The over-simplified picture presented above would imply $T_\alpha=T_\mathrm{K}$ \citep{Field} but this is assuming that there are no spectral distortions in the Ly~$\alpha$ spectrum. Models for $x_\alpha$ and $T_\alpha$ have improved over the years. Some of the obvious corrections would be the following. Firstly, due to scattering the specific intensity goes down in the vicinity of Ly~$\alpha$ resonance and so does $x_\alpha$. Moreover, the energy exchange between \ion{H}{i} and Ly~$\alpha$ photons causes $T_\alpha$ not to relax to $T_\mathrm{K}$ but to somewhere between $T_\mathrm{K}$ and $T_\mathrm{s}$ \citepalias{Chen}. We explore both these aspects later in this section. Further details such as fine and hyperfine structure of Ly~$\alpha$ and frequency dependence of spin-flip probability have been considered in \citet[hereafter \citetalias{Hirata}]{Hirata}, but here we do not consider them as they are relevant for very low kinetic temperatures.

To evaluate $x_\alpha$ and $T_\alpha$ we need the specific intensity of Ly~$\alpha$ photons denoted by $J(\nu)$. We define it in terms of number (\emph{not} energy) per unit proper area per unit proper time per unit frequency per unit solid angle. The specific intensity is obtained by solving the equation of radiative transfer under the Fokker--Planck approximation \citep{Rybicki1994}. However, it is generally not possible to find the analytical expressions for $x_\alpha$ and $T_\alpha$. The results of \citetalias{Chen} and \citetalias{Hirata} are quite accurate but they rely on numerical approach and iterative techniques. 

For our work we use the analytical solution for the spectrum around a general resonance line under the `wing approximation' by \citet{Grachev} or more specifically the work of \citet{Chuzhoy} for Ly~$\alpha$ photons. In the wing approximation the Voigt line profile is approximated by the `wings' of the Lorentzian line. \citetalias{FP06} have examined the validity of this in detail by comparing results using full line profile with that using wing approximation. Their conclusion is that accuracy is not sacrificed when using the latter (but may be important at extremely low kinetic temperatures).

We now present the spectrum of Ly~$\alpha$ radiation. For mathematical convenience the linearity of equation of radiative transfer, under the Fokker--Planck approximation, allows us to split the solution into two parts: spectrum of continuum photons $J_\mathrm{c}(\nu)$ and that of injected photons $J_\mathrm{i}(\nu)$, so that $J=J_\mathrm{c}+J_\mathrm{i}$. Physically, the difference between the two lies in their origin. The photons released by the stars between the Ly~$\alpha$ and Ly~$\beta$ which redshift and ultimately give Ly~$\alpha$ photons are called continuum photons. The photons between Ly~$\gamma$ and Ly~$\infty$ will redshift to Ly~$\gamma$ or other higher Lyman series lines. These higher Lyman lines can decay to Ly~$\alpha$ photons via radiative cascade. These comprise the injected photons. We do not include photons between Ly~$\beta$ and Ly~$\gamma$ because they redshift to Ly~$\beta$ which never end up in Ly~$\alpha$ resonance. This is because selection rule tells that a 3p configuration decays to 1s or 2s and the latter always undergoes a two-photon emission \citepalias{Hirata}. 

Let the undisturbed background Ly~$\alpha$ specific intensity far from the resonance line be $J_\alpha=J_\alpha(z)$ (we will discuss its calculation in Section~\ref{Jalpha}) and for now we assume that it is same for both, the continuum and injected photons. The specific intensity of continuum photons is\footnote{Note the misprint in equation (10) of \citet{Chuzhoy}: inside the integral the argument of exponent should have a $z^3$ instead of $x^3$.} \citepalias{FP06}
\begin{multline}
J_\mathrm{c}(x)=\frac{2\pi J_\alpha}{a\tau_{\alpha}}\exp\left[-2\eta x-\frac{2\pi x^3}{3a\tau_{\alpha}}\right]\\\times\int_{-\infty}^{x}y^2\exp\left[2\eta y+\frac{2\pi y^3}{3a\tau_{\alpha}}\right]\,\ud y\,,\label{Jc}
\end{multline}
and for injected photons
\begin{equation}
J_\mathrm{i}(x)=J_\mathrm{i}(0)\exp\left[-2\eta x-\frac{2\pi x^3}{3a\tau_{\alpha}}\right]\text{ for }x\geqslant0\label{Ji}
\end{equation}
whereas for $x<0$, $J_\mathrm{i}(x)$ is the same as $J_\mathrm{c}(x)$. The changed variable, Voigt parameter\footnote{We find a discrepancy in the Voigt parameter, $a$, by \citetalias{Chen}. The denominator should have 4 instead of 8.}, Doppler width and the recoil parameter are given by
\begin{align}
&x=\frac{\nu-\nu_\alpha}{\Delta\nu_\mathrm{D}}\,,\label{x}\\
&a=\frac{A_\alpha}{4\pi\Delta\nu_\mathrm{D}}\,,\\
&\Delta\nu_\mathrm{D}=\nu_\alpha\sqrt{\frac{2k_\mathrm{B}T_\mathrm{K}}{m_\text{H}c^2}}\,\quad\mathrm{and},\\
&\eta=\frac{h/\lambda_\alpha}{\sqrt{2m_{\text{H}}k_{\mathrm{B}}T_{\mathrm{K}}}}\,,\label{eta}
\end{align}
respectively. Here $A_\alpha=\SI{6.25e8}{\hertz}$ is the Einstein spontaneous emission coefficient of Ly~$\alpha$ transition, $m_\text{H}$ is the mass of recoiling atom (here hydrogen), $\lambda_\alpha(\nu_\alpha)$ is the wavelength (frequency) of the Ly~$\alpha$ photon, $c$ is the speed of light and the Ly~$\alpha$ optical depth \citep{GP} is given by
\begin{equation}
\tau_{\alpha}=\frac{3\gamma_\alpha\lambda_\alpha^3n_\text{H}x_{\ion{H}{i}}}{2H}\,,    
\end{equation}
where $\gamma_\alpha$ is the half width at half maximum of Ly~$\alpha$ resonance line given by \citepalias{Hirata}
\begin{equation}
\gamma_\alpha=\frac{e^2\mathcal{F}_\alpha\nu_\alpha^2}{6m_\text{e}c^3\varepsilon_0}=\SI{50}{\mega\hertz}\,,\label{HWHM}
\end{equation}
where $e$ is the charge of electron, $m_{\text{e}}$ is its mass, $\varepsilon_0$ is the permittivity of vacuum and $\mathcal{F}_\alpha=0.4182$ is the oscillator strength of Ly~$\alpha$ resonance. Usually the functions $J(x)$ are written in terms of Sobolev parameter $\gamma_\mathrm{S}$ related to $\tau_\alpha$ as $\gamma_\mathrm{S}=\tau_\alpha^{-1}$.

\begin{figure*}
\centering
\includegraphics[width=0.9\textwidth]{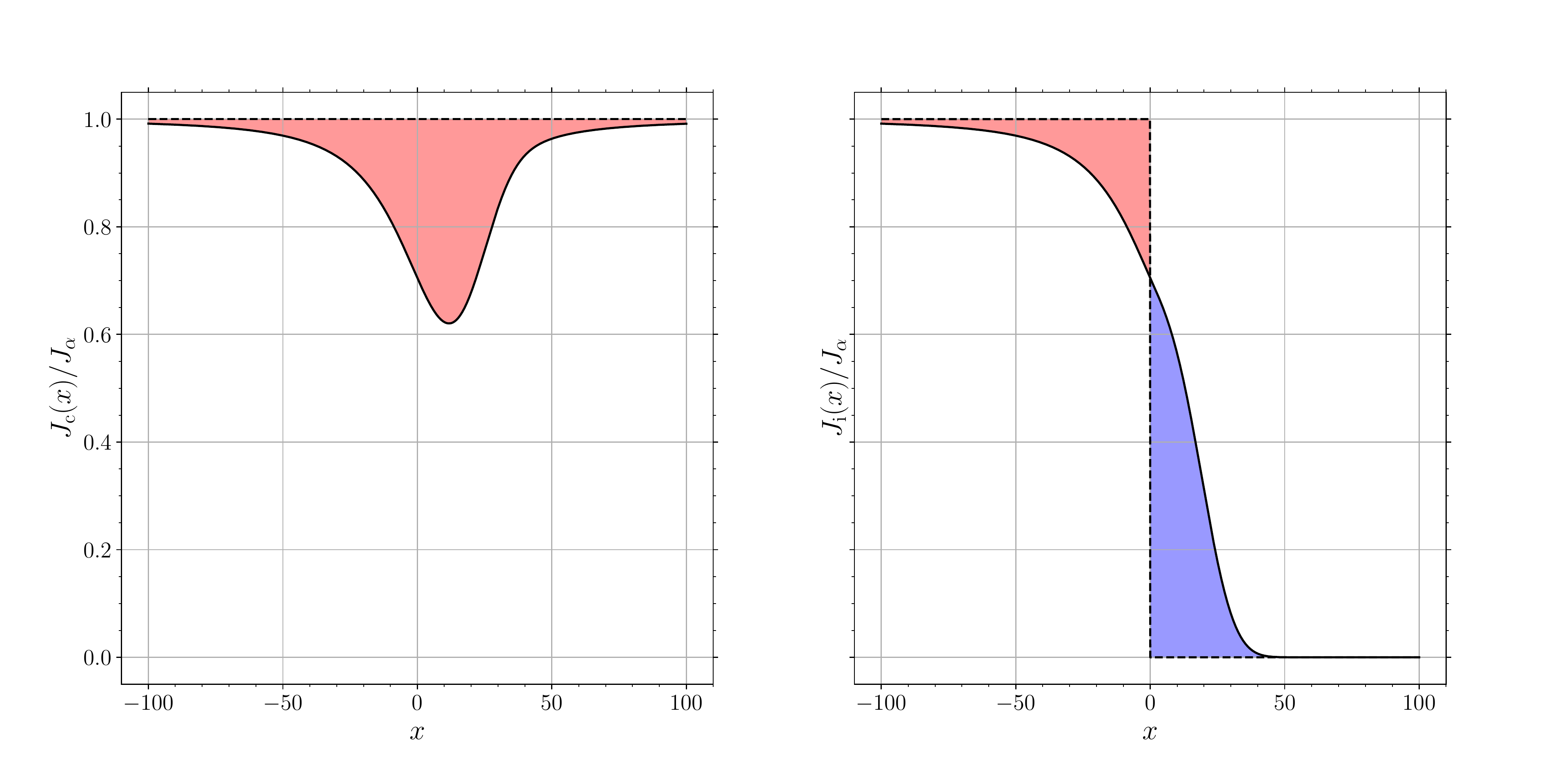}
\caption{Specific intensity of continuum and injected Ly~$\alpha$ photons normalised to the same background intensity. These curves are plotted at $(z, x_\text{e},T_\mathrm{K})\sim(22,0,\SI{10}{\kelvin})$. The left panel shows continuum photons (equation~\ref{Jc}). The area of the shaded region is $I_\mathrm{c}\approx20.11$. The right panel shows injected photons (equation~\ref{Ji}). The blue shaded area represents cooling $(I_\mathrm{i}^{\text{cool}}\approx-13.07)$, while the red one represents heating ($I_\mathrm{i}^{\text{heat}}\approx7.32$). The asymmetry exists because of the extra cosmological redshift due to the expanding Universe. See text for more details. The dashed lines correspond to the case for no scattering of Ly~$\alpha$ photons, or the infinite temperature limit.}\label{Lya}
\end{figure*}

Figure~\ref{Lya} shows the spectra $J_{\mathrm{c}}$ and $J_{\mathrm{i}}$. The asymmetry in the spectrum can be explained qualitatively as follows. Because of the Doppler effect, when the source (stars) and detector (atom) are moving towards each other the apparent frequency measured by the detector increases and when moving away it decreases. When a hydrogen atom in the IGM is hit with radiation, it will selectively absorb photons of frequency $\nu_\alpha$ as measured in its rest frame. The resulting spectrum would of course be a Lorentzian in the rest frame. If the Universe was static then this spectrum would still be symmetric when transformed to the lab frame, although it will be more broadened due to the Doppler broadening. However, in an expanding Universe even the sources are also moving and away. Thus, the whole spectrum would come out to be shifted to a higher frequency in order to compensate for this added cosmological redshift. This explains the asymmetry in the spectra shown in Figure~\ref{Lya}.

With the specific intensity function at hand we can now discuss the coupling $x_\alpha$ and colour temperature $T_\alpha$. The probability that a Ly~$\alpha$ photon will bring an \ion{H}{i} from the upper hyperfine state to a lower one (indirectly, via WF effect) is approximately $4/27$ \citep[\citetalias{Hirata}]{Meiksin00} so that if $P_\alpha$ is the total rate of Ly~$\alpha$ photon scattering per hydrogen atom then
\begin{equation}
P_{10}=\frac{4}{27}P_\alpha\,.\label{P10}
\end{equation}
The quantity $P_\alpha$ is given by
\begin{equation}
P_\alpha=\frac{\pi e^2\mathcal{F}_\alpha}{m_{\text{e}}\varepsilon_0c} \int_{-\infty}^{\infty} J(\nu)\phi_\alpha(\nu) \ud\nu\,,\label{pa}
\end{equation}
where $\phi_\alpha(\nu)$ is the normalised Ly~$\alpha$ line profile. In the wing approximation it looks like (expressing in terms of dimensionless frequency $x$)
\begin{equation}
\phi_\alpha(x)\approx\frac{a}{\pi x^2}\,.
\end{equation}
Using equations \eqref{P10} and \eqref{pa} in equation \eqref{xa} we get
\begin{align}
x_\alpha&=\frac{4\pi e^2\mathcal{F}_\alpha }{27A_{21} m_{\text{e}}\varepsilon_0c}\frac{T_*}{T_\gamma}J_\alpha \int_{-\infty}^{\infty} \frac{J(\nu)}{J_\alpha}\phi_\alpha(\nu) \ud\nu\\
&\equiv S\frac{J_\alpha}{J_0}\,,\label{xalpha}
\end{align}
where 
\begin{equation}
S=\int_{-\infty}^{\infty} \frac{J(\nu)}{J_\alpha}\phi_\alpha(\nu) \ud\nu\,,  \label{SC}
\end{equation}
is called the scattering correction and the quantity $J_0$ is given by
\begin{align}
J_{0}&=\frac{27A_{21} m_{\text{e}}\varepsilon_0c}{4\pi e^2\mathcal{F}_\alpha}\frac{T_\gamma}{T_*}\\
&\approx\num{5.54e-8}(1+z)\,\si{\per\metre\squared\per\second\per\hertz\per\steradian}\,,
\end{align}
where we used $T_\gamma=T_0(1+z)$. An accurate formula for $S$ can be obtained by assuming that line profile is sharply peaked at $\nu=\nu_\alpha$ or equivalently $x=0$ so that $S\approx J_\mathrm{i}(x=0)/J_\alpha$. The trick to find $J_\mathrm{i}(0)$ and hence $S$ is to exploit the continuity of specific intensity of injected photons at $x=0$.  This gives
\begin{equation}
  J_\mathrm{i}(0)=\frac{2\pi J_\alpha}{a\tau_{\alpha}}\int_{-\infty}^{0}y^2\exp\left[2\eta y+\frac{2\pi y^3}{3a\tau_{\alpha}}\right]\,\ud y\,,
\end{equation}
so that
\begin{equation}
S=\frac{2\pi}{a\tau_{\alpha}}\int_{-\infty}^{0}y^2\exp\left[2\eta y+\frac{2\pi y^3}{3a\tau_{\alpha}}\right]\,\ud y\,.\label{S1}
\end{equation}
Different expressions in closed form can be found in literature for $S$ (cf.~\citealt{Chuzhoy} and \citetalias{FP06}). We derive yet another expression (Appendix~\ref{AppA}), which is more condensed, given by
\begin{equation}
S=1-{}_{\phantom{1}3}F_0(1/3,2/3,1;0;-\xi_1)\,,\label{S}
\end{equation}
where 
\begin{equation}
\xi_1=\frac{9\pi}{4a\tau_\alpha\eta^3}\,,
\end{equation}
and${}_{\phantom{1}3}F_0$ is the $(3,0)$-hypergeometric function \citep[Chap. 18,][]{ARFKEN}. A typical value of $S$ would be $\sim0.7$ for $(z,x_{\text{e}},T_\mathrm{K})\sim(22,0,\SI{10}{\kelvin})$.

The formal definition of the colour temperature is \citep{Rybicki2006}
\begin{equation}
T_\alpha^{-1}=-\frac{k_\mathrm{B}}{h}\frac{\ud\ln \mathcal{N}(\nu)}{\ud\nu}\,,\label{talpha}  
\end{equation}
where $\mathcal{N}(\nu)=c^2J(\nu)/(2\nu^2)$ is the photon occupation number\footnote{The photon occupation number should not be confused with the specific number density of photons often denoted by $n(\nu)$ or $n_\nu$. They are related as $n(\nu)=8\pi\nu^2\mathcal{N}(\nu)/c^3$.}. Clearly, $T_\alpha$ is a frequency-dependent quantity but to obtain a number out of the right hand side of equation~\eqref{talpha}, it should be averaged over the line profile $\phi_\alpha$ \citep{Meiksin}. Alternatively, we may evaluate it at $\nu=\nu_\alpha$ (approximation same as the one used for evaluating $S$). Since the resulting $T_\alpha$s differ by a small amount, we may take the latter approach. The final expression we will use is \citep{Chuzhoy}
\begin{equation}
T_\alpha=T_\mathrm{s}\left(\frac{T_\mathrm{K}+T_{\text{se}}}{T_\mathrm{s}+T_{\text{se}}}\right)\,, \label{Tc}
\end{equation}
where 
\begin{equation}
T_{\text{se}}=\left(\frac{\nu_{21}}{\nu_\alpha}\right)^2\frac{m_{\text{H}}c^2}{9k_\mathrm{B}}\approx\SI{0.4}{\kelvin}\,.
\end{equation}
The smallness of $T_{\text{se}}$ -- which captures the correction due to spin exchange -- is indicative of the fact that for temperatures above a few kelvin the argument $T_\alpha=T_\mathrm{K}$ of \citet{Field} is quite accurate. The spin exchange correction also modifies the recoil parameter and the Sobolev parameter but here we are neglecting those effects since they are important only at extremely low temperatures, typically $T_\mathrm{K}\lesssim\SI{1}{\kelvin}$ (\citetalias{Hirata}, \citetalias{FP06}). 

We can eliminate $T_\alpha$ from equation~\eqref{Ts} and write the spin temperature as
\begin{equation}
T_\mathrm{s}^{-1}=\frac{T_\gamma^{-1}+x_\mathrm{K}T_\mathrm{K}^{-1}+x_\alpha(T_\mathrm{K}+T_{\text{se}})^{-1}}{1+x_\mathrm{K}+x_\alpha T_\mathrm{K}(T_\mathrm{K}+T_{\text{se}})^{-1}}\,.
\end{equation}
The above may be approximated as \citep{BARKANA_2016}
\begin{equation}
T_\mathrm{s}^{-1}=\frac{T_\gamma^{-1}+(x_\mathrm{K}+x_\alpha)T_\mathrm{K}^{-1}}{1+x_\mathrm{K}+x_\alpha}\,.
\end{equation}

\subsection{\texorpdfstring{Heating by Ly~$\alpha$ Photons}{Heating by Ly~α Photons}}
We now consider the role of Ly~$\alpha$ photons in heating the IGM. We can qualitatively understand it as follows. The continuum photons descend from higher frequency and are preferentially scattered off by atoms moving away from them. As a result they continually lose energy and cause heating. Stated differently, in the absence of scattering the spectral distortion would redshift away. However, in steady state the photons would lose energy to atoms continuously. In the case of injected photons, some of them are scattered off to the blue side by the atoms moving in the opposite direction, which create a cooling effect, while the remaining are scattered off to the red side and produce a heating effect. The net effect by the continuum and injected photons is quite small.

The heat supplied by the continuum photons per unit time per unit proper volume is given by (\citetalias{Chen}\footnote{There is a typo in equation~(10) of \citetalias{Chen}. The factor $\Delta\nu_\mathrm{D}$ should not be present. Their equations~(17) \& (18), however, are correct.})
\begin{equation}
q_\mathrm{c}=\frac{4\pi}{c} Hh\int_{-\infty}^{\infty}\nu(J_\alpha-J_\mathrm{c})\,\ud\nu\,.
\end{equation}
We may approximate $\nu=\nu_\alpha$ throughout the integral. Changing the variable to $x=(\nu-\nu_\alpha)/\Delta\nu_\mathrm{D}$ we can write
\begin{equation}
\frac{2q_\mathrm{c}}{3n_{\mathrm{b}}k_\mathrm{B}H}=\frac{8\pi}{3}\frac{h}{k_\mathrm{B}\lambda_\alpha}\frac{J_\alpha \Delta\nu_\mathrm{D}}{n_{\mathrm{b}}(z)} I_\mathrm{c}\,,\label{Qc}
\end{equation}
where
\begin{equation}
I_\mathrm{c}=\int_{-\infty}^{\infty}\left[1-\frac{J_\mathrm{c}(x)}{J_\alpha}\right]\,\ud x\,.\label{Ic}
\end{equation}
We have explicitly shown the $z$ dependence of $n_{\mathrm{b}}$ on the right hand side to remind ourselves that both $J_\alpha$ and $n_{\mathrm{b}}$ are in proper units. Graphically, $I_\mathrm{c}$ is an area between the undisturbed Ly~$\alpha$ spectrum (which is a flat line) and a scattered one. See, for e.g., the left panel of Figure~\ref{Lya} plotted at $(z,x_{\text{e}},T_\mathrm{K})\approx(22,\num{2.19e-4},\SI{10}{\kelvin})$ in which the red shaded area is $I_\mathrm{c}\approx20.11$. The expression for $I_\mathrm{c}$ can be written in a closed form as \citepalias{FP06}
\begin{equation}
I_\mathrm{c}=\eta(2\pi^4a^2\tau_\alpha^2)^{1/3}\left[\text{Ai}^2(-\xi_2)+\text{Bi}^2(-\xi_2)\right]\,,
\end{equation}
where
\begin{equation}
\xi_2=\eta\left(\frac{4a\tau_\alpha}{\pi}\right)^{1/3}\,,
\end{equation}
Ai($x$) and Bi($x$) represent the Airy function of first and second kind, respectively \citep{airy}.

We can also write an equation similar to equation~\eqref{Qc} for injected photons by changing subscript `c' to `i'. However, $I_\mathrm{i}$ is defined as
\begin{equation}
I_\mathrm{i}=\int_{-\infty}^{0}\left[1-\frac{J_\mathrm{c}(x)}{J_\alpha}\right]\,\ud x-\int_{0}^{\infty} \frac{J_\mathrm{i}(x)}{J_\alpha}\,\ud x\,.
\end{equation}
The first integral in $I_\mathrm{i}$ can only be simplified to
\begin{multline}
\int_{-\infty}^{0}\left[1-\frac{J_\mathrm{c}(x)}{J_\alpha}\right]\,\ud x=\eta\sqrt{\frac{a\tau_\alpha}{2}}\int_{0}^{\infty}\left[\exp\left(-2\eta y-\frac{\pi y^3}{6a\tau_\alpha}\right)\right.\\\times\left.\text{erfc}\sqrt{\frac{\pi y^3}{2a\tau_\alpha}}\frac{\ud y}{\sqrt{y}}\,\right],
\end{multline}
where erfc($x$) represents the complementary error function \citep[Chap. 13,][]{ARFKEN}. The second integral is
\begin{equation}
\int_{0}^{\infty} \frac{J_\mathrm{i}(x)}{J_\alpha}\,\ud x=\frac{J_\mathrm{i}(0)}{J_\alpha}\int_{0}^{\infty}\exp\left[-2\eta x-\frac{2\pi x^3}{3a\tau_{\alpha}}\right]\ud x\,.
\end{equation}
We already approximated $S$ by $J_\mathrm{i}(0)/{J_\alpha}$. As for the integral, we split the exponential into two and do integration by parts to get
\begin{multline}
-\frac{1}{2\eta}\left.\exp\left[-2\eta x-\frac{2\pi x^3}{3a\tau_{\alpha}}\right]\right|_{0}^{\infty}\\+\frac{1}{2\eta}\frac{2\pi}{a\tau_{\alpha}}\int_{0}^{\infty}x^2\exp\left[-2\eta x-\frac{2\pi x^3}{3a\tau_{\alpha}}\right]\,\ud x
\end{multline}
\begin{equation}
=\frac{1}{2\eta}-\frac{S}{2\eta}\,,
\end{equation}
where in the second term we changed the variable to $x=-y$ to get the integral for $S$ as in equation~\eqref{S1}. So finally we get,
\begin{multline}
I_\mathrm{i}=\eta\sqrt{\frac{a\tau_\alpha}{2}}\int_{0}^{\infty}\exp\left[-2\eta y-\frac{\pi y^3}{6a\tau_\alpha}\right]\text{erfc}\sqrt{\frac{\pi y^3}{2a\tau_\alpha}}\frac{\ud y}{\sqrt{y}}\\-\frac{S(1-S)}{2\eta}\,.\label{cool}
\end{multline}
with $S$ given by equation~\eqref{S}. Generally, the effect of injected photons is to cool the IGM except at extremely low gas kinetic temperatures $(T_\mathrm{K}\lesssim\SI{1}{\kelvin}$) but such low temperatures are not realised at redshifts of our interest. For the example shown in the right panel of Figure~\ref{Lya}, $I_\mathrm{i}\approx-5.75$.

In the preceding discussion we assumed that the background intensity of continuum and injected photons is the same but this is not true in general. However, we can easily correct for this by specifying the ratio $r=J_\alpha^\mathrm{i}/J_\alpha^\mathrm{c}$, which depends on the surface temperature of the source. We take
\begin{equation}
r=\frac{J_\alpha^\mathrm{i}}{J_\alpha^\mathrm{c}}=0.1\,,
\end{equation}
appropriate for Population-II (Pop-II) stars \citep[cf. \citetalias{Chen}, in which $r= 1$]{Chuzhoy_2007}. Combining the contributions from continuum and injected photons, the total heating rate by Ly~$\alpha$ photons is
\begin{equation}
\frac{2q_{\alpha}}{3n_{\mathrm{b}}k_\mathrm{B}H}=\frac{8\pi}{3}\frac{h}{k_\mathrm{B}\lambda_\alpha}\frac{J_\alpha(z) \Delta\nu_\mathrm{D}}{n_{\mathrm{b}}(z)} \left(I_\mathrm{c}+\frac{J_\alpha^\mathrm{i}}{J_\alpha^\mathrm{c}} I_\mathrm{i}\right)\,.\label{qalpha}
\end{equation}
We ignore the small recoil heating contribution from deuterium atom \citep{Chuzhoy_2007}.

\begin{figure}
\centering
\includegraphics[width=1\linewidth]{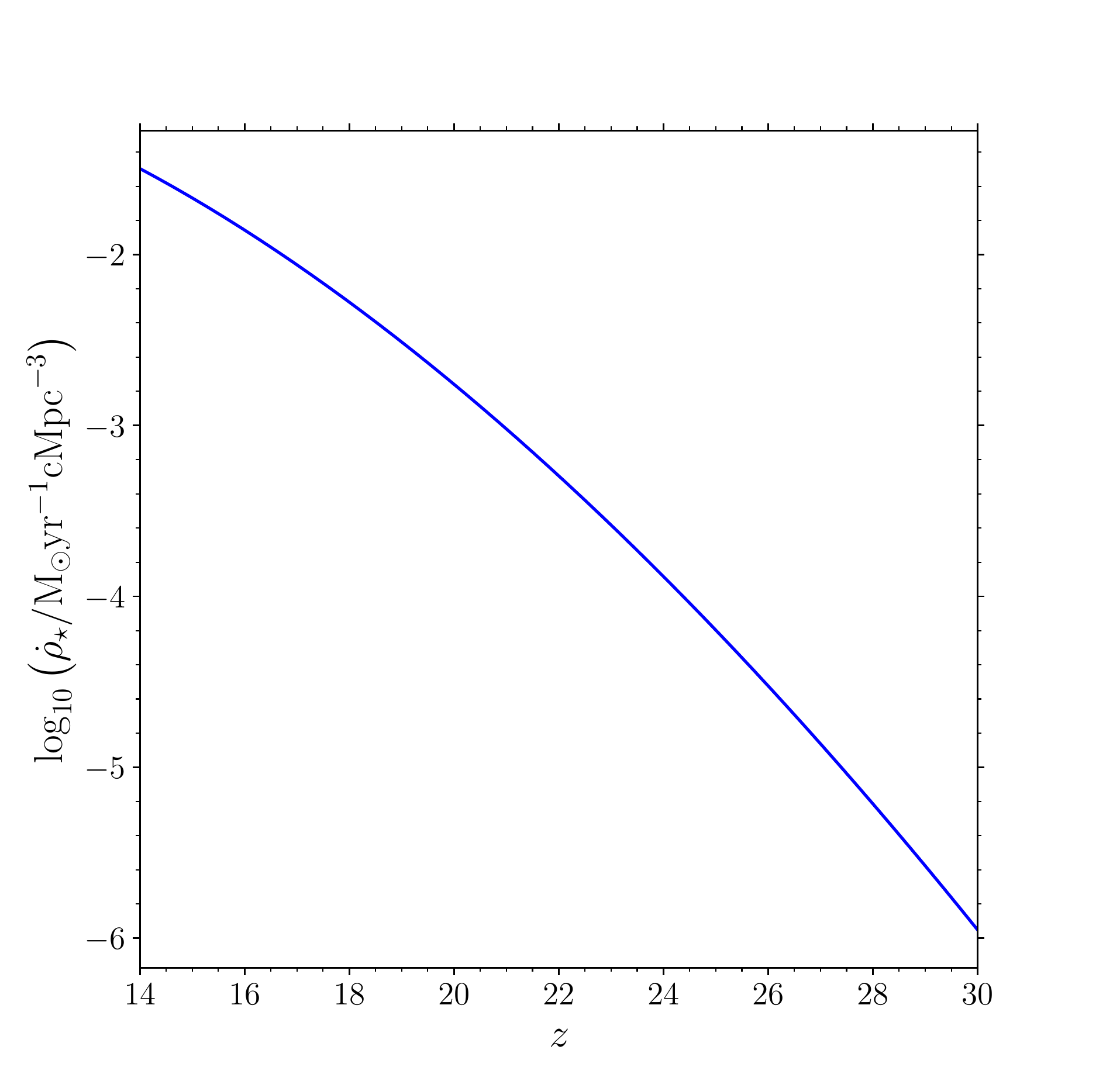}
\caption{The comoving star formation rate density for a star formation efficiency of $f_\star=0.1$ and halo virial temperatures of $T_{\text{vir}}\geqslant\SI{e4}{\kelvin}$.}\label{SFRD_plot}
\end{figure}

\subsection{\texorpdfstring{The Background Ly~$\alpha$ Specific Intensity}{The Background Lyα Specific Intensity}}\label{Jalpha}
To calculate $J_\alpha(z)$ we need the comoving UV emissivity $\epsilon_{\text{UV}}(E,z)$. The comoving emissivity is defined as the number of photons emitted per unit comoving volume per unit proper time per unit energy at redshift $z$ and energy $E$. To model it we assume that it is proportional to the star formation rate density (SFRD) and spectral energy distribution (SED) \citep{Barkana}. More precisely, Emissivity = (the number of UV photons emitted per unit energy at $E$ per baryon in the stars)$\times$(number of baryons accumulating in the stars per unit time per unit comoving volume at $z$), i.e.,
\begin{equation}
\epsilon_{\text{UV}}(E,z)=\epsilon_\mathrm{b}(E)\frac{\dot{\rho}_\star(z)}{m_\mathrm{b}}\,,\label{epsilon}
\end{equation}
where $\epsilon_\mathrm{b}(E)$ is the SED, defined as the number of photons emitted per baryon per unit energy and $m_\mathrm{b}$ is the average baryon mass. For the redshift range of our interest we can accurately write $m_\mathrm{b}=1.22m_{\text{H}}$ \citep{Haimoud}. 

The SED depends on the source or the type of star but it is generally a broken power law, i.e., $\epsilon_\mathrm{b}(E)\propto E^{s-1}$ where the index $s$ can be different between different Lyman lines. We assume the model of Pop-II stars which emit $N_{\alpha\beta}=6520$ photons per baryon between Ly~$\alpha$ and Ly~$\beta$ with index $s=0.14$. Between Ly~$\beta$ and Ly~$\infty$ they emit $N_{\beta\infty}=3170$ photons per baryon, so that the total is $N_{\alpha\infty}=9690$ \citep{Barkana}. To find the proportionality constants and the index for the latter case we used the normalisation and continuity at $E_\beta=\SI{12.09}{\electronvolt}$, which is the energy of the Ly~$\beta$ line. We derive the final expression for $\epsilon_\mathrm{b}$ in $\si{\per\electronvolt}$
\begin{equation}
\epsilon_\mathrm{b}(E)=
\begin{cases}
2902.91\, \hat{E}^{-0.86}&\text{if }E\in[E_\alpha,E_\beta]\\
1303.34\, \hat{E}^{-7.66}&\text{if }E\in(E_\beta,E_\infty]\,,
\end{cases}
\end{equation}
where $\hat{E}=E/E_\infty$, $E_\alpha=\SI{10.2}{\electronvolt}$ and $E_\infty=\SI{13.6}{\electronvolt}$ are the energies corresponding to Ly~$\alpha$ and Ly~$\infty$ transition, respectively.

The comoving SFRD is represented by $\dot{\rho}_\star(z)$, and is measured in mass per unit time per unit comoving volume. We assume it is proportional to the rate at which baryons collapse into dark matter haloes. Assuming only the haloes of virial temperatures $(T_{\text{vir}})$ above $\SI{e4}{\kelvin}$ contribute, their number at a given redshift can be determined by the \citet{Press} formalism. Thus,
\begin{equation}
\dot{\rho}_\star(z)=-(1+z)\bar{\rho}_\mathrm{b}^0f_\star H(z)\frac{\ud F_{\text{coll}}(z)}{\ud z}\,,
\end{equation}
where 
\begin{equation}
\bar{\rho}_\mathrm{b}^0=\frac{3H_0^2}{8\pi G}\Omega_\mathrm{b}\,,
\end{equation}
is the mean cosmic baryon mass density measured today, $f_\star(=0.1)$ is the star formation efficiency, defined as the fraction of baryons converted into stars in the haloes. We denote the fraction of baryons that have collapsed into dark matter haloes by $F_{\text{coll}}$ given by the following expression \citep{Bar}
\begin{equation}
F_{\text{coll}}(z)=\mathrm{erfc}\left[\frac{\delta_{\mathrm{crit}}(z)}{\sqrt{2}\sigma(m_{\mathrm{min}})}\right]\,,
\end{equation}
where $\delta_{\mathrm{crit}}$ is the linear critical density of collapse and $\sigma^2$ is the variance in smoothed density field. The minimum halo mass for star formation is
\begin{equation}
m_{\mathrm{min}}=10^8\frac{1}{\sqrt{\Omega_{\mathrm{m}}}}\left(\frac{H_0}{100}\right)^{-1}\mathrm{M}_{\odot}\left[\frac{10}{1+z}\frac{0.6}{\mu}\frac{\mathrm{min}(T_{\text{vir}})}{\num{1.98e4}}\right]^{3/2}\,,
\end{equation}
where $H_0$ is the Hubble's constant measured today in units of $\SI{1}{\kilo\metre\per\second\per\mega\parsec}$. For min($T_{\text{vir}})=\SI{e4}{\kelvin}$ and $\mu=1.22$ the above simplifies to
\begin{equation}
m_{\mathrm{min}}=\num{3.91e8}\frac{1}{\sqrt{\Omega_{\mathrm{m}}}}\left(\frac{H_0}{100}\right)^{-1}\mathrm{M}_{\odot}(1+z)^{-3/2}\,.
\end{equation}
One may calculate the quantity $\delta_{\mathrm{crit}}(z)/\sigma(m_{\mathrm{min}})$ using the \textsc{colossus} code\footnote{\url{https://bitbucket.org/bdiemer/colossus/src/master/}} \citep{Colossus}. As an example, for our cosmological parameters we get $F_{\text{coll}}(z=0)\approx0.735$. See Figure~\ref{SFRD_plot} for a plot of the SFRD as a function of redshift.

It is a good approximation to account for the effect of higher Lyman series (Ly~$n$) photons only in the total Ly~$\alpha$ intensity, since analogous WF effect of Ly~$n$ or the direct heating by them is negligible (\citetalias{FP06}, however, see \citet{Meiksin_10} for a different point of view). We can now write $J_\alpha$ as
\begin{equation}
J_\alpha(z)=\frac{c}{4\pi}(1+z)^2\sum_{n=2}^{23}P_n\int_z^{z_{\text{max}}}\frac{\epsilon_{\text{UV}}(E_n',z')}{H(z')}\,\ud z'\,,\label{J}
\end{equation}
where the $n^{\text{th}}$ term in the sum accounts for the finite probability $P_n$ with which a photon in the upper Lyman lines will redshift to Ly~$\alpha$ wavelength. The values of $P_n$ are computed in an iterative fashion using the selection rule and the decay rates. The detailed procedure and table of values can be found in \citetalias{Hirata} or \citet{PF06}. The redshifted energy of $n^{\text{th}}$ Lyman series line is given by
\begin{equation}
E_n'=E_n\frac{1+z'}{1+z}\,, 
\end{equation}
where $E_n$ is the frequency of the photon released in transition from $n^{\text{th}}$ state to ground state
\begin{equation}
E_n=13.6\left(1-\frac{1}{n^2}\right)\,\si{\electronvolt}\,.
\end{equation}
The maximum redshift from which this photon could have been received is given by
\begin{equation}
1+z_{\text{max}}=\frac{E_{n+1}}{E_{n}}(1+z)=\frac{1-(1+n)^{-2}}{1-n^{-2}}(1+z)\,.    
\end{equation}
In writing equation~\eqref{J} we implicitly assumed that Ly~$\alpha$ photons stream freely across the IGM and reach the line centre at the same distance from the source. However, in reality Ly~$\alpha$ radiation would suffer multiple scatterings with hydrogen atoms and set up a stable background at different distances from the source. Formally, there should be a distance-dependent transmission probability factor to account for this effect \citep{Zheng, Semelin, Naoz, higgins, Higgins_12, Reis20}. We are ignoring these complications here.

\begin{figure*}
\centering
\includegraphics[width=1\textwidth]{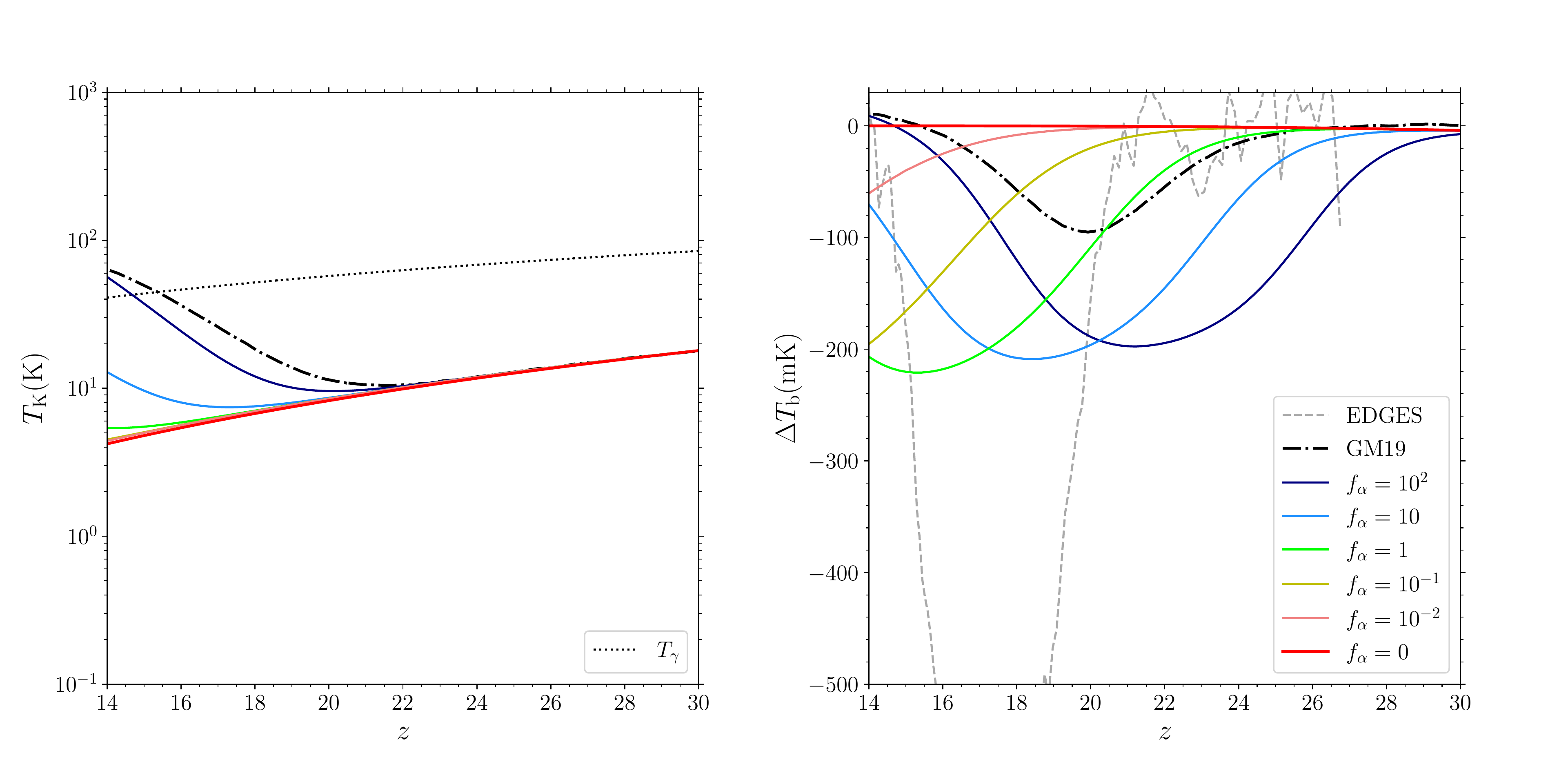}
\caption{Left panel: Evolution of the gas kinetic temperature $T_\mathrm{K}$. The black dotted line is the CMB temperature $T_\gamma=2.73(1+z)$. The red solid line $(f_\alpha=0)$ corresponds to the adiabatic cooling in which case $T_\mathrm{K}\approx0.02(1+z)^2$ \citep{Scott}. Right panel: the corresponding differential brightness temperature or the 21-cm signal using equation~\eqref{DeltaT}. The grey dashed line is the EDGES detection of the global 21-cm signal. In both panels the thick black dash dotted line shows the fiducial model from \citetalias{Ghara} for comparison.}\label{Fig3}
\end{figure*}

\section{Results and Analysis}\label{RA}

We now consider the magnitude of Ly~$\alpha$ heating expected under our model assumptions. In order to gauge the strength of the Ly~$\alpha$ background we parametrize the SED using a scaling parameter $f_\alpha$ (cf.~\citetalias{Ghara}).  We introduce this parameter by writing $\epsilon_{\mathrm{b}}$ as $f_\alpha \epsilon_{\mathrm{b}}$ so that
\begin{equation}
\epsilon_{\text{UV}}(E,z)=f_\alpha\epsilon_\mathrm{b}(E)\frac{\dot{\rho}_\star(z)}{m_\mathrm{b}}\,,
\end{equation}
Note how the effect of this change propagates 
\begin{equation*}
f_\alpha\rightarrow \epsilon_\mathrm{b}(E)\rightarrow \epsilon(E,z)\rightarrow J_\alpha\rightarrow x_\alpha, q_\alpha\,.
\end{equation*}
Thus, both the coupling and heating are affected as we change $f_\alpha$. We will consider six values for it: $f_\alpha=\{0,10^{-2},10^{-1},1,10,10^2\}$.

Figure~\ref{Fig3} shows our result.  In the left panel we show the variation of gas kinetic temperature for different values of $f_\alpha$. The corresponding plots of differential brightness are shown in the right panel of the same figure. Note how the timing and the depth of the absorption feature changes as $f_\alpha$ is changed. The case $f_\alpha=0$ corresponds to a Universe where there is no Ly~$\alpha$ radiation so that $q_\alpha,x_\alpha=0$. In such a scenario the matter temperature just falls as $(1+z)^2$ as expected for pure adiabatic cooling and the 21-cm signal would be practically a null signal (shown in thick red). This is because the collisional coupling $x_\mathrm{K}$ in this era is very small and hence the spin temperature is close to CMB temperature. In the right panel of Figure~\ref{Fig3} we have also shown the global 21-cm signal reported by the EDGES collaboration \citep{Bowman} in grey dashed line.

Our important finding is that for identical astrophysical assumptions, we find a much reduced Ly~$\alpha$ heating than recent literature. As an example, for $f_\alpha=1$, Ly~$\alpha$ heating in \citetalias{Ghara} becomes significant at $z\sim 22$ while in our case it remains subdominant until $z\sim16$. At $z=14$, the IGM temperature in our model is an order of magnitude lower ($T_\mathrm{K}\sim\SI{6}{\kelvin}$) than that in \citetalias{Ghara}.  This obviously affects the 21-cm signal absorption feature which in our model occurs at $z\sim16$ and has an amplitude of $\SI{-220}{\milli\kelvin}$. 

\citetalias{Ghara} claim that larger values of $f_\alpha$ are ruled out, however, our results say otherwise. If we want the signal to be more negative then from equation~\eqref{DeltaT} we can say that $T_\mathrm{s}$ should be as small as possible. But from equation~\eqref{Ts} the theoretical minimum of $T_\mathrm{s}$ can only be the lowest of all quantities being averaged, which here is $T_\mathrm{K}$. For this to happen the weight factor $x_\alpha$ should be as high as possible since $x_\mathrm{K}\approx0$. For e.g., when $f_\alpha=1$ then at $z=22$ we get $(x_{\text{e}},T_\mathrm{K})=(\num{2.19e-4},\SI{10.14}{\kelvin})$ for which $x_\alpha\approx0.24$ whereas $x_\mathrm{K}\approx\num{3.2e-3}$. Thus, in view of the EDGES signal \citep{Bowman} we can conclude that the optimum strength of Ly~$\alpha$ background for Pop-II stars would be characterised by
\begin{equation}
    1<f_\alpha<10\,,
\end{equation}
and that too without any excess cooling models such as a phenomenological cooling \citep{Mirocha} or a physically motivated cooling \citep[such as][]{Bar18}. If we include one of those then we could push $f_\alpha$ to even higher values to get stronger coupling at a negligible cost of extra Ly~$\alpha$ heating. We therefore conclude that the EDGES measurement does not rule out significant build-up of a Ly~$\alpha$ background at cosmic dawn. (See Appendix~\ref{AppB} for a more detailed comparison of our work with the literature.)

We have not varied the other possibly free parameters in this study such as $r=J_\alpha^\mathrm{i}/J_\alpha^\mathrm{c}$, $f_\star$ and $T_{\text{vir}}$. The latter two are degenerate with $f_\alpha$. \citet{Mirocha} argue that $f_\star$ should take higher values, which is consistent with our results. The ratio $r$, however, is an interesting parameter. In this study we chose its value to be 0.1 but if it increases, then the heating effect by continuum photons would get cancelled by the increased cooling by injected photons (see equation~\ref{qalpha}) with no decrement in $x_\alpha$. For the 21-cm signal this means that the absorption feature becomes deeper, which is more favourable for us, again, keeping in mind the EDGES result.  

\section{Conclusions}\label{Con}

In this work we saw how Ly~$\alpha$ photons affect the global 21-cm cosmological signal. Their scattering by the neutral hydrogen atoms in the intergalactic medium couples the spin temperature to the gas kinetic temperature. Also, the recoil induced in the hydrogen atom as a result of this scattering heats up the IGM. We used an analytical expression for the spectrum of Ly~$\alpha$ radiation obtained by the wing and the Fokker--Planck approximations. Using this expression and exploiting the sharply-peaked nature of the line profile it was possible to write an analytical expression for the scattering correction and for the heating rate due to the continuum photons, although this was not possible for the heating rate due to the injected photons. We derived two new expressions in this work. An expression for the scattering correction $S$ (equation~\ref{S}) and an expression for the pure cooling part of the injected photons (equation~\ref{cool}). We did not consider the effect of deuterium anywhere.  We also did not consider any direct scattering/heating effect due to higher Lyman-series photons but only accounted for them via radiative cascade in writing the total undisturbed background radiation.

We assumed Population-II type stars as the source of Ly~$\alpha$ radiation. In order to study the effect of the Ly~$\alpha$ background strength on the gas kinetic temperature and hence the 21-cm differential brightness we varied the amplitude of spectral energy distribution by four orders of magnitude while retaining its power-law shape. Our key finding in this study is that a strong Ly~$\alpha$ background is necessary in order to produce a strong absorption signal.  This is contrary to recent conclusions in the literature \citep{Ghara}. For our fiducial model $f_\alpha=1$ (the parameter which quantifies the intensity of Ly~$\alpha$ background) we find that the IGM remains significantly colder than the CMB down to at least $z=14$. As we discuss in Appendix~\ref{AppB}, while our result disagrees with \citetalias{Ghara}, our Ly~$\alpha$ heating formalism agrees with that presented by \citet{Chuzhoy_2007} and \citet{Meiksin} in a different form. We can thus say quite generally that Ly~$\alpha$ heating will be small at cosmic dawn unless our astrophysical assumptions regarding Ly~$\alpha$ production are dramatically changed. 

Beyond Ly~$\alpha$, the relative strengths of X-ray and ultraviolet radiation are also quite uncertain. Modelling each effect brings along its own set of free parameters. The resultant parameter space can be explored to find the model which best fits the EDGES signal as well as satisfies other cosmological constraints. Some studies which have taken this approach include \citet{Cohen17, Greig18, Monsalve_2019, Cohen20}. Some recent papers have provided new insights on old physics. \citet{Meiksin20} considered an enhanced Ly~$\alpha$ radiation from Population-III stars which can create a cooling effect if reddened by winds internal to the haloes. Similarly, \citet{mebane} considered effects of X-ray and radio emission from Pop-III stars on the 21-cm signal.

We have not attempted to match our signal with the EDGES measurement by means of exotic cooling or excess radio background models. We will do so in future and work with a larger redshift range which encompasses not only Ly~$\alpha$ heating but also includes the important effects such as photoheating by X-rays \citep{Mesinger12, Mesinger, Christian_2013}, shock heating \citep{F04}, and reionization \citep{Haard}.

\section*{Acknowledgements}

It is a pleasure to acknowledge discussions with members of the Radio Experiment for the Analysis of Cosmic Hydrogen (REACH) collaboration. GK gratefully acknowledges support by the Max Planck Society via a partner group grant. We also thank James Bolton, Anastasia Fialkov, Raghunath Ghara and Itamar Reis for their comments.

\section*{Data availability}

No new data were generated or analysed in support of this research.


\bibliographystyle{mnras}
\bibliography{Biblo} 



\appendix

\section{A derivation for the scattering correction, \textit{S}}\label{AppA}
Here we derive equation~\eqref{S}. Starting from equation~\eqref{S1} we have
\begin{equation}
S=\frac{2\pi}{a\tau_{\alpha}}\int_{-\infty}^{0}y^2\exp\left[2\eta y+\frac{2\pi y^3}{3a\tau_{\alpha}}\right]\,\ud y\,.
\end{equation}
Make a change of variable: $u=-2\eta y$ and set 
\begin{equation}
\xi_1=\frac{9\pi}{4a\tau_\alpha\eta^3}\,,
\end{equation}
to get
\begin{equation}
S=\int_{0}^{\infty}\left[\frac{\xi_1u^2}{9}\ue^{-\xi_1(u/3)^3}\right] \ue^{-u}\,\ud u\,.
\end{equation}
Integration by parts gives
\begin{equation}
S=1-\int_{0}^{\infty}\ue^{-\xi_1(u/3)^3} \ue^{-u}\,\ud u\,.
\end{equation}
Make a Taylor's series expansion of the first exponential to get
\begin{equation}
S=1-\sum_{n=0}^\infty\frac{1}{3^{3n}}\left[\int_{0}^{\infty}u^{3n} \ue^{-u}\,\ud u\right] \frac{(-\xi_1)^n}{n!}\,,
\end{equation}
where `!' represents the regular factorial. The integral is just the gamma function \citep[Chap. 13,][]{ARFKEN}, hence
\begin{equation}
S=1-\sum_{n=0}^\infty\frac{(3n)!}{3^{3n}} \frac{(-\xi_1)^n}{n!}\,.
\end{equation}
In terms of the Pochhammer symbol or the rising factorial \citep[Chap. 18,][]{ARFKEN} the above expression can be rewritten as
\begin{equation}
S=1-\sum_{n=0}^\infty(1/3)_n(2/3)_n(1)_n \frac{(-\xi_1)^n}{n!}\,.
\end{equation}
By the definition of generalised hypergeometric function
\begin{equation}
S=1-{}_{\phantom{1}3}F_0(1/3,2/3,1;0;-\xi_1)\,.
\end{equation}

\section{Comparison with the literature}\label{AppB}

The purpose of this appendix is to compare and contrast various expressions available in the literature for the volumetric heating rate of the intergalactic medium (IGM) by the Lyman-$\alpha$ (Ly~$\alpha$) photons. This appendix is organized as follows. In Sections~\ref{CM04} and \ref{M06}, we derive the volumetric Ly~$\alpha$ heating rate expression used by \citet[\citetalias{Chen}]{Chen} and \citet[\citetalias{Meiksin}]{Meiksin}, respectively. In Sections~\ref{CS07} and \ref{GM19}, we analyse the volumetric Ly~$\alpha$ heating rate expression used by \citet[\citetalias{Chuzhoy_2007}]{Chuzhoy_2007} and \citet[\citetalias{Ghara}]{Ghara}, respectively.

\subsection{\texorpdfstring{Ly~$\alpha$ heating rate given by \citetalias{Chen}}{Ly~α heating rate given by CM04}}\label{CM04}
In this section we derive the volumetric heating rate due to Ly~$\alpha$ photons given by \citetalias{Chen}. Their expression agrees with the expression that we have used in this work. 

To begin, we follow \citet{Rybicki2006} to obtain the energy exchange between Ly~$\alpha$ radiation and \ion{H}{i} atoms. For this, they wrote down the time derivative of energy density of radiation and used the radiative transfer equation. We will not repeat those steps here and start from their equation~(37). Thus, the volumetric energy transfer rate to \ion{H}{i} from Ly~$\alpha$ radiation is
\begin{equation}
-\frac{\partial U}{\partial t}\equiv q=4\pi k_\mathrm{B}T_\mathrm{K}\frac{h\nu_\alpha^2}{m_\text{H}c^2}\chi\int \phi_\alpha(\nu)\left(\frac{\partial {J}}{\partial \nu}+\frac{h{J}}{k_\mathrm{B}T_\mathrm{K}}\right)\ud\nu\,,\label{Eq1}
\end{equation}
where the symbols have the usual meaning (see Section~\ref{TM}). For equation~\eqref{Eq1}, the usual assumption of line profile being sharply peaked is made so that $\nu^2$ goes outside the integral and becomes $\nu_\alpha^2$.

The variable $\chi$ has different names and conventions in literature. We define it as follows \citep{Rybicki2006, Rybicki1994}
\begin{equation}
\chi=\frac{h\nu_\alpha}{4\pi}n_1B_{12}=n_1\frac{e^2\mathcal{F}_\alpha}{4m_\text{e}\varepsilon_0 c}\equiv n_1\sigma_\alpha\,,\label{Eq2}
\end{equation}
where $n_1$ is the population density of the ground state of hydrogen and $B_{12}$ is the Einstein coefficient of stimulated absorption. For other symbols see the description of equation~\eqref{HWHM}. In writing $\chi$, we have ignored stimulated emission since the upper-level population (first excited state) is quite negligible because of high spontaneous emission coefficient of Ly~$\alpha$ line. So most of the hydrogen atoms tend to stay in the ground state. For the same reason we may make the following approximation
\begin{equation}
n_1\approx n_{\ion{H}{i}}\,,\label{Eq3}
\end{equation}
where $n_{\ion{H}{i}}$ is the neutral hydrogen number density. The dimensions of $\chi$ are $[\mathrm{L}^{-1}\mathrm{T}^{-1}]$.

Our first step in simplification is to convert the quantities in dimensionless form. Using equations~\eqref{x} and \eqref{eta}, equation~\eqref{Eq1} becomes
\begin{align}
q &=4\pi k_\mathrm{B}T_\mathrm{K}\frac{h\nu_\alpha^2}{m_\text{H}c^2}\chi\frac{1}{\Delta\nu_\mathrm{D}}\int \left(\frac{\partial J}{\partial x}+\frac{h\Delta\nu_\mathrm{D} J}{k_\mathrm{B}T_\mathrm{K}}\right)\phi_\alpha(x)\ud x\\
&=4\pi h\frac{\Delta\nu_{\mathrm{D}}^2}{2}\chi \frac{1}{\Delta\nu_\mathrm{D}}\int \left(\frac{\partial J}{\partial x}+2\eta J\right)\phi_\alpha(x)\ud x\\
&=4\pi h\Delta\nu_\mathrm{D} \chi\int_{-\infty}^{\infty} \frac{\phi_\alpha}{2}\frac{\partial J}{\partial x}+\eta\phi_\alpha J\,\ud x\,,\label{Eq4}
\end{align}
where we have used $\phi_\alpha(\nu)\ud \nu=\phi_\alpha(x)\ud x$ and $J(\nu)\to J(x)\equiv J$ is now a function of $x$. Note that the dimensions of $J$ remain intact but $\phi_\alpha(x)$ is now dimensionless.

The heating rate $q$ has the dimensions of energy per unit time per unit volume. Using integration by parts for the integral in equation~\eqref{Eq4} we get (for the moment, ignore the factors outside the integration)
\begin{equation}
\left.x\left(\frac{\phi_\alpha}{2}\frac{\partial J}{\partial x}+\eta\phi_\alpha J\right)\right|_{-\infty}^{\infty}-\int_{-\infty}^{\infty} x\frac{\partial}{\partial x}\left(\frac{\phi_\alpha}{2}\frac{\partial J}{\partial x}+\eta\phi_\alpha J\right)\ud x\,,
\end{equation}
The first term is 0, because $\phi_{\alpha}(x)$ is a rapidly decaying function and thus
\begin{equation}
q=-4\pi h\Delta\nu_\mathrm{D} \chi\int_{-\infty}^{\infty} x\frac{\partial}{\partial x}\left(\frac{\phi_\alpha}{2}\frac{\partial J}{\partial x}+\eta\phi_\alpha J\right)\ud x\,.\label{Eq5}
\end{equation}
The above equation is reduced to the form of equation~(A4) of \citetalias{Chen}\footnote{There are two errors in equation~(A4) of \citetalias{Chen}. There is a minus sign missing and instead of a $\Delta\nu_\mathrm{D}$ factor, there should be $\Delta\nu_\mathrm{D}^2$. However, their main equations~(17) \& (18) are correct.}.

Now let us see how we can get the heating rate due to continuum photons (equation~\ref{Qc}). The specific intensity, $J_{\mathrm{c}}(x)$, of continuum photons satisfy the following Fokker--Planck equation in steady state \citepalias{Chen}
\begin{equation}
\frac{\partial}{\partial x}\left(\frac{\phi_\alpha}{2}\frac{\partial J_{\mathrm{c}}}{\partial x}+\eta\phi_\alpha J_{\mathrm{c}}\right)+\gamma_\mathrm{S}\frac{\partial J_{\mathrm{c}}}{\partial x}=0\,,\label{Eq6}
\end{equation}
where $\gamma_\mathrm{S}$ is the Sobolev parameter, which captures the effect of the expansion of the Universe and is given by \citep{Rybicki1994}
\begin{equation}
\gamma_\mathrm{S}=\frac{H}{\chi\lambda_\alpha}\,.\label{Eq7}
\end{equation}
Using equation~\eqref{Eq6} in equation~\eqref{Eq5} we get
\begin{equation}
q_\mathrm{c}=4\pi h\Delta\nu_\mathrm{D} \chi \gamma_{\mathrm{S}}\int_{-\infty}^{\infty}x\frac{\partial J_{\mathrm{c}}}{\partial x} \ud x\,.\label{Eq8}
\end{equation}
Integrating by parts gives us (as before, considering the integral only)
\begin{equation}
\left.(xJ_{\mathrm{c}})\right|_{-\infty}^{\infty}-\int_{-\infty}^{\infty}J_{\mathrm{c}}\, \ud x\,.
\end{equation}
If the undisturbed and undistorted spectrum far from the resonance line is denoted by $J_\alpha$, i.e. $\lim_{x\to\pm\infty}J_{\mathrm{c}}(x)=J_\alpha$, then the above can be written as
\begin{align*}
&\,J_\alpha\int_{-\infty}^{\infty}\ud x-\int_{-\infty}^{\infty}J_{\mathrm{c}}\, \ud x\\
=&\,\int_{-\infty}^{\infty}(J_\alpha-J_{\mathrm{c}})\,\ud x\\
=&\,J_\alpha I_\mathrm{c}\,,
\end{align*}
where $I_\mathrm{c}$ was defined in equation~\eqref{Ic}. Inserting the above in equation~\eqref{Eq8} we get
\begin{equation}
q_\mathrm{c}=4\pi h\Delta\nu_\mathrm{D} \frac{H}{\lambda_\alpha} J_\alpha I_\mathrm{c}\,.\label{Eq9}
\end{equation}
where we used equation~\eqref{Eq7} for the definition of Sobolev parameter. In dimensions of temperature
\begin{equation}
\frac{2q_\mathrm{c}}{3n_{\mathrm{b}}k_\mathrm{B}H}=\frac{8\pi}{3}\frac{h}{k_\mathrm{B}\lambda_\alpha}\frac{J_\alpha \Delta\nu_\mathrm{D}}{n_{\mathrm{b}}(z)} I_\mathrm{c}\,.\label{Eq10}
\end{equation}
Equation~\eqref{Eq10} is now in exactly the form of equation~\eqref{Qc}. The above calculation can be easily extended to include the injected photons to arrive at
\begin{equation}
q_\alpha=4\pi h\Delta\nu_\mathrm{D} \frac{H}{\lambda_\alpha} J_\alpha \left(I_\mathrm{c}+\frac{J_\alpha^\mathrm{i}}{J_\alpha^\mathrm{c}} I_\mathrm{i}\right)\,,\label{Eq11}
\end{equation}
which when expressed in units of temperature gives equation~\eqref{qalpha}.

This type of formalism is used by \citetalias{Chen} and \citet{FP06}.

\subsection{\texorpdfstring{Ly~$\alpha$ heating rate given by \citetalias{Meiksin}}{Ly~α heating rate given by M06}}\label{M06}
In this section we derive the volumetric heating rate due to Ly~$\alpha$ photons given by \citetalias{Meiksin} \citep[cf.][]{Rybicki2006}. It is a seemingly different, albeit equivalent, form of $q_\alpha$. The starting point remains the same as before, namely equation~\eqref{Eq1}. Let us rewrite it as
\begin{multline}
q_\alpha=4\pi k_\mathrm{B}T_\mathrm{K}\frac{h\nu_\alpha^2}{m_\text{H}c^2}\chi \frac{h}{k_\mathrm{B}T_\mathrm{K}}\int {J}(\nu)\phi_\alpha(\nu)\\\times\left(1+T_\mathrm{K}\frac{k_\mathrm{B}}{h}\frac{\partial\ln {J(\nu)}}{\partial\nu}\right)\ud\nu\,.\label{Eq12}
\end{multline}
But the definition of colour temperature as employed by \citetalias{Chen} is
\begin{equation}
T_\alpha^{-1}=-\left.\frac{k_\mathrm{B}}{h}\frac{\partial\ln {J(\nu)}}{\partial\nu}\right|_{\nu\approx\nu_\alpha}\,,\label{Eq13}
\end{equation}
so that equation~\eqref{Eq12} becomes
\begin{align}
q_\alpha&=4\pi k_\mathrm{B}T_\mathrm{K}\frac{h\nu_\alpha^2}{m_\text{H}c^2}\chi \frac{h}{k_\mathrm{B}T_\mathrm{K}}\int {J}(\nu)\phi_\alpha(\nu)\left(1-\frac{T_\mathrm{K}}{T_\alpha}\right)\ud\nu\\
&=4\pi \frac{(h\nu_\alpha)^2}{m_\text{H}c^2}n_{\text{H}}\sigma_\alpha \int {J}(\nu)\phi_\alpha(\nu)\ud\nu\left(1-\frac{T_\mathrm{K}}{T_\alpha}\right)\,,\label{Eq14}
\end{align}
where we used equation~\eqref{Eq2} with approximation \ref{Eq3} (and a further approximation, $n_{\ion{H}{i}}\approx n_{\mathrm{H}}$, for a nearly neutral universe). Using the expression of total scattering rate of Ly~$\alpha$ photons $P_\alpha$ from equation \eqref{pa} we get
\begin{equation}
q_\alpha=\frac{(h\nu_\alpha)^2}{m_\text{H}c^2}n_{\text{H}}P_\alpha\left(1-\frac{T_\mathrm{K}}{T_\alpha}\right)\,.\label{Eq15}
\end{equation}
This type of formula is used by \citetalias{Meiksin} and \citet{Meiksin20}.

Note some obvious differences between the two types. In this formalism the contribution of continuum and injected photons is included implicitly through the value of colour temperature. This is because $J(\nu)$ employed to calculate $T_\alpha$ in equation~\eqref{Eq13} is the sum of intensities of continuum and injected photons. However, this formulation makes it explicitly clear as to why Ly~$\alpha$ heating is so small; because of the closeness of $T_\mathrm{K}$ and $T_\alpha$. In standard cosmological scenario we have \citep{Meiksin20}
\begin{equation}
\left|1-\frac{T_\mathrm{K}}{T_\alpha}\right|\sim\num{e-4}\textrm{--}\num{e-3}\,.\label{Eq16}
\end{equation}

\subsection{\texorpdfstring{Ly~$\alpha$ heating rate given by \citetalias{Chuzhoy_2007}}{Ly~α heating rate given by CS07}}\label{CS07}
In Sections~\ref{CM04} and \ref{M06} we have established that equations~\eqref{Eq11} and \eqref{Eq15} are correct and equivalent descriptions of the heating rate due to scattering of Ly~$\alpha$ photons.

In this section we explore an apparently different formula for Ly~$\alpha$ heating by \citetalias{Chuzhoy_2007}. According to them, the total energy gained by \ion{H}{i} from each Ly~$\alpha$ photon is (in our notation style and using $\nu$ instead of $x$)
\begin{equation}
\Delta E_{\mathrm{Tot}}=\int \frac{J(\nu)}{J_\alpha}\Delta E(\nu)\phi_\alpha(\nu)\,\ud\nu\,\label{Eq17}
\end{equation}
where\footnote{The second term in square brackets of equation~(4) of \citetalias{Chuzhoy_2007} should have a $\Delta\nu_{\mathrm{D}}$ in the denominator, unless by $\phi'(x)$ they meant $\ud \phi(x)/\ud\nu$. However, mathematically speaking, this notation is incorrect.}
\begin{equation}
\Delta E(\nu)=\frac{(h\nu_\alpha)^2}{m_{\mathrm{H}}c^2}\left[1-\frac{k_\mathrm{B}T_\mathrm{K}}{h}\frac{\phi'_\alpha(\nu)}{\phi_\alpha(\nu)}\right]\,,\label{Eq18}
\end{equation}
and $\phi'(\nu)=\ud\phi_\alpha(\nu)/\ud\nu$. Combining equations~\eqref{Eq17} and \eqref{Eq18} we get
\begin{equation}
\Delta E_{\mathrm{Tot}}=\frac{(h\nu_\alpha)^2}{m_{\mathrm{H}}c^2}\int \frac{J(\nu)}{J_\alpha}\left[1-\frac{k_\mathrm{B}T_\mathrm{K}}{h}\frac{\phi'_\alpha(\nu)}{\phi_\alpha(\nu)}\right]\phi_\alpha(\nu)\,\ud\nu\,.
\end{equation}
Separating out the terms,
\begin{multline}
\Delta E_{\mathrm{Tot}}=\frac{(h\nu_\alpha)^2}{m_{\mathrm{H}}c^2}\left\lbrace\int \frac{J(\nu)}{J_\alpha}\phi_\alpha(\nu)\,\ud\nu\right.\\-\left.\frac{k_\mathrm{B}T_\mathrm{K}}{hJ_\alpha}\int\phi'_\alpha(\nu)J(\nu)\,\ud\nu\right\rbrace\,.\label{Eq19}
\end{multline}
Consider only the second integral from the above equation. Apply integration by parts to it as follows
\begin{align}
&\ \int_{-\infty}^{\infty}\phi'_\alpha(\nu)J(\nu)\,\ud\nu\\
&=\left.\phi_\alpha(\nu)J(\nu)\right|_{-\infty}^{\infty}-\int_{-\infty}^{\infty}\phi_\alpha(\nu)J'(\nu)\,\ud\nu\,,
\end{align}
where $J'(\nu)=\ud J(\nu)/\ud\nu$. The first term in the above goes to 0, since $\lim_{\nu\to\pm\infty} \phi_\alpha(\nu)=0$. Inserting the remaining expression into equation~\eqref{Eq19}, $\Delta E_{\mathrm{Tot}}$ becomes
\begin{multline}
\Delta E_{\mathrm{Tot}}=\frac{(h\nu_\alpha)^2}{m_{\mathrm{H}}c^2}\left\lbrace\int \frac{J(\nu)}{J_\alpha}\phi_\alpha(\nu)\,\ud\nu\right.\\+\left.\frac{k_\mathrm{B}T_\mathrm{K}}{hJ_\alpha}\int\phi_\alpha(\nu)J'(\nu)\,\ud\nu\right\rbrace
\end{multline}
\begin{equation}
\qquad\quad =\frac{(h\nu_\alpha)^2}{m_{\mathrm{H}}c^2}\int \frac{J(\nu)}{J_\alpha}\left[1+\frac{k_\mathrm{B}T_\mathrm{K}}{h}\frac{J'(\nu)}{J(\nu)}\right]\phi_\alpha(\nu)\,\ud\nu\,.
\end{equation}
Using the definition of colour temperature from equation~\eqref{Eq13} we get
\begin{equation}
\Delta E_{\mathrm{Tot}}=\frac{(h\nu_\alpha)^2}{m_\text{H}c^2}\left(1-\frac{T_\mathrm{K}}{T_\alpha}\right)\int\frac{J(\nu)}{J_\alpha}\phi_\alpha(\nu)\,\ud\nu\,.\label{Eq20}
\end{equation}
Let the number of photons that cross Ly~$\alpha$ resonance per H atom per unit time be $\dot{N}_\alpha$ \citepalias{Chuzhoy_2007}. It is given by\footnote{This expression of $\dot{N}_\alpha$ was missing in \citetalias{Chuzhoy_2007} but we infer this based on its definition provided and so that the final formula for $q_\alpha^{\mathrm{CS07}}$ turns out to be correct.}
\begin{equation}
\dot{N}_\alpha=4\pi\sigma_\alpha J_\alpha\,,\label{nalpha}
\end{equation}
where $J_\alpha$ is the background Ly~$\alpha$ specific intensity (see Section~\ref{Jalpha}). Multiplying equation~\eqref{Eq20} by $\dot{N}_\alpha$ and $n_{\mathrm{H}}$, we get the volumetric heating rate as
\begin{equation}
q_\alpha^{\mathrm{CS07}}=\frac{(h\nu_\alpha)^2}{m_\text{H}c^2}n_{\text{H}}\left(1-\frac{T_\mathrm{K}}{T_\alpha}\right)4\pi\sigma_\alpha\int J(\nu)\phi_\alpha(\nu)\,\ud\nu\,.
\end{equation}
Finally, expressing the above in terms of $P_\alpha$ using equation~\eqref{pa} we get
\begin{equation}
q_\alpha^{\mathrm{CS07}}=\frac{(h\nu_\alpha)^2}{m_\text{H}c^2}n_{\text{H}}P_\alpha\left(1-\frac{T_\mathrm{K}}{T_\alpha}\right)\,.\label{Eq21}
\end{equation}
Thus, the formula given by \citetalias{Chuzhoy_2007} has reduced to the form of equation~\eqref{Eq15}. A similar formalism was used by \citet{Ciardi}.

\subsection{\texorpdfstring{Ly~$\alpha$ heating rate given by \citetalias{Ghara}}{Ly~α heating rate given by GM19}}\label{GM19}

In this section we analyse yet another formula for Ly~$\alpha$ heating rate used by \citetalias{Ghara} in their work. Although it has appearance similar to that of \citetalias{Chuzhoy_2007} but the individual terms in the formula are different from that of \citetalias{Chuzhoy_2007}. It is written as
\begin{equation}
q^{\mathrm{GM19}}_\alpha=n_{\text{H}}\dot{N}_\alpha^{\mathrm{GM19}} \left(\Delta E_\mathrm{c}+\frac{J_\alpha^\mathrm{i}}{J_\alpha^\mathrm{c}}\Delta E_\mathrm{i}\right)\,,\label{eqn:gm19rate}
\end{equation}
where the individual terms in the above expression are\footnote{Since these details were missing in their paper, we thank Raghunath Ghara for providing them in private communication.}
\begin{align}
\dot{N}_\alpha^{\mathrm{GM19}}&=\frac{27x_\alpha}{4}\frac{T_\gamma A_{21}}{T_*}\,,\\
\Delta E_\mathrm{c}&=\frac{(h\nu_\alpha)^2}{m_\text{H}c^2}\int_{-\infty}^{\infty}\left[1-\frac{J_{\mathrm{c}}(x)}{J_\alpha}\right]\phi_\alpha(x)\, \ud x\,,
\end{align}
and
\begin{multline}
\Delta E_\mathrm{i}=\frac{(h\nu_\alpha)^2}{m_\text{H}c^2}\left(\int_{-\infty}^{0}\left[1-\frac{J_\mathrm{c}(x)}{J_\alpha}\right]\phi_{\alpha}(x)\,\ud x\right.\\-\left.\int_{0}^{\infty} \frac{J_\mathrm{i}(x)}{J_\alpha}\phi_\alpha(x)\,\ud x\right)\,.
\end{multline}
It is clear from the above expressions that they cannot be derived from first principles.

Having given the formula, we now analyse it in some detail. For clarity, let us rewrite $\dot{N}_\alpha^{\mathrm{GM19}}$ in our notation using equations~\eqref{xa} and \eqref{P10}. We get
\begin{equation}
\dot{N}_\alpha^{\mathrm{GM19}}=P_\alpha=4\pi\sigma_\alpha \int J(\nu)\phi_\alpha(\nu)\,\ud\nu\,. \label{eqn:ndot}
\end{equation}
Thus, $\dot{N}_\alpha^{\mathrm{GM19}}$ is just the total scattering rate of Ly~$\alpha$ photons. Further let us approximate the integrals of $\Delta E$ as follows
\begin{equation}
\int_{-\infty}^{\infty}\left[1-\frac{J_{\mathrm{c}}(x)}{J_\alpha}\right]\phi_\alpha(x)\, \ud x\approx 1-S\,,\label{eqn:dE}
\end{equation}
assuming that $\phi_\alpha(x)$ is sharply peaked at $x=0$. This gives
\begin{equation}
\Delta E_\mathrm{c}\approx \frac{(h\nu_\alpha)^2}{m_\text{H}c^2}(1-S)\,.\label{DE_c}
\end{equation}
Similarly, 
\begin{equation}
\Delta E_\mathrm{i}\approx \frac{(h\nu_\alpha)^2}{m_\text{H}c^2}\left(\frac{1}{2}-S\right)\,,\label{DE_i}
\end{equation}
Combining equations~\eqref{eqn:ndot}, \eqref{DE_c} and \eqref{DE_i} for $r=0.1$, we get the final expression of total volumetric heating rate by Ly~$\alpha$ photons used by \citetalias{Ghara} as follows
\begin{equation}
q_{\alpha}^{\mathrm{GM19}}\approx1.1\frac{(h\nu_\alpha)^2}{m_\text{H}c^2}n_{\text{H}}P_\alpha(0.95-S)\,.\label{qalphagm}
\end{equation}
We now show that $q_{\alpha}^{\mathrm{GM19}}$ is nearly two orders of magnitude higher than the correct version of $q_\alpha$.
Dividing equation~\eqref{qalphagm} by our equation~\eqref{Eq11} we get
\begin{equation}
\frac{q_{\alpha}^{\mathrm{GM19}}}{q_\alpha}=\frac{1.1h^2\nu_\alpha^2n_{\text{H}}P_\alpha(0.95-S)/(m_\text{H}c^2)}{4\pi h\Delta\nu_\mathrm{D} (H/\lambda_\alpha) J_\alpha (I_\mathrm{c}+0.1I_{\mathrm{i}})}\,,
\end{equation}
which can be written as 
\begin{equation}
\frac{q_\alpha^{\mathrm{GM19}}}{q_\alpha}=1.1\frac{S(0.95-S)\eta n_{\text{H}}\lambda_\alpha\sigma_\alpha}{H I}\,.
\end{equation}
where for simplicity $I=(I_\mathrm{c}+0.1I_{\mathrm{i}})$. Putting in the numbers and trends for $I$ and $S$ we find
\begin{equation}
\frac{q_\alpha^{\mathrm{GM19}}}{q_\alpha}\sim 15(1+z)^{0.6}\,.
\end{equation}
Using equation~\eqref{qalphagm} directly, or our equation~\eqref{Eq11} multiplied by the above factor, reproduces \citetalias{Ghara}'s results. A typical value of this factor would be, say at $z=23$, $q_\alpha^{\mathrm{GM19}}/q_\alpha\approx101$. We can thus quite generally say that the Ly~$\alpha$ heating rate computed by \citetalias{Ghara} is erroneously higher than its correct value by a factor of $\sim\mathcal{O}(10^2)$.  This explains the difference between our conclusions about constraints from EDGES on the Ly~$\alpha$ background at Cosmic Dawn. (Note that given the fundamental difference between the heating rates used by \citetalias{Ghara} and \citetalias{Chuzhoy_2007}, the `agreement' between the results of \citetalias{Ghara} and \citetalias{Chuzhoy_2007} asserted by \citetalias{Ghara} is probably accidental and is most likely due to a drastic difference in the Ly~$\alpha$ emissivity models used in these two papers.)


\bsp	
\label{lastpage}
\end{document}